\documentstyle[12pt,epsf,epsfig,wrapfig]{article}

\everymath{\displaystyle}
\newcommand{\mb}[1]{\mbox{\boldmath$\bf#1$}}
\begin{document}

\begin{center}

{\bfseries Particle and spin motion in polarized media}

\vskip 5mm

A.J. Silenko

\vskip 5mm

{\small

{\it

Institute of Nuclear Problems, Belarusian State University, Minsk, Belarus

}

{\it E-mail: silenko@inp.minsk.by}}

\end{center}

\vskip 5mm

\begin{abstract}

Quantum mechanical equations of motion are obtained for particles
and spin in media with polarized electrons in the presence of
external fields. The motion of electrons and their spins is
governed by the exchange interaction, while the motion of
positrons and their spins is governed by the annihilation
interaction. The equations obtained describe the motion of
particles and spin in both magnetic and nonmagnetic media. The
evolution of positronium spin in polarized media is investigated.
Media with polarized electrons can be used for polarization of
positronium beams.

\end{abstract}

\newpage

%\section {Introduction}
\noindent{\bfseries 1 ~~ Introduction}

\vskip 5mm

The quantum mechanical description of the motion of particles and
spin in matter is a very important problem. The classical theory
of motion of particles and spin has been developed in great detail
(see \cite{TB,YB})). A quantum mechanical equation of motion of
relativistic particles in an electromagnetic field was derived by
Derbenev and Kondratenko \cite{DK}. The motion of the spin of
relativistic particles in an electromagnetic field is described by
the Bargmann--Michel--Telegdi (BMT) equation \cite{BMT}. The
Lagrangian with an allowance for terms quadratic in spin was
obtained in \cite{PK,PS}. The corresponding equation of spin
motion was presented in \cite{YP}. The interaction between
polarized particles and polarized matter was analyzed in
\cite{Bar1,JETP}.

In the present paper, we find quantum mechanical equations of
motion of particles and spin for relativistic particles with
arbitrary spin that move in media with polarized electrons in the
presence of external fields. The system of units $\hbar=c=1$ is
used.

\vskip 5mm

%\section {Hamiltonian for particles in polarized media}
\noindent{\bfseries 2 ~~ Hamiltonian for particles in polarized
media}

\vskip 5mm

For particles with arbitrary spin, the Hamiltonian can be derived
with the use of the interaction Lagrangian ${\cal L}$, obtained in
\cite{PK,PS}. This Lagrangian contains terms that are linear
(${\cal L}_1$) and quadratic (${\cal L}_2$) in spin:
\begin{equation} \begin{array}{c}
{\cal L}={\cal L}_1+{\cal L}_2, ~~~ {\cal
L}_1=\frac{e}{2m}\left\{\left(g-2+\frac{2}{\gamma}\right)(\mb
S\cdot \mb B)-(g-2)\frac{\gamma}{\gamma+1}(\mb S\cdot\mb v)(\mb
v\cdot\mb B)+ \right. \\ \left.
\left(g-2+\frac{2}{\gamma+1}\right)(\mb S\cdot[\mb E\times\mb
v])\right\},
\\ {\cal L}_2=\frac{Q}{2s(2s-1)}\left[(\mb S\cdot\nabla)-
\frac{\gamma}{\gamma+1}(\mb S\cdot\mb v)(\mb v\cdot\nabla)\right]
\left[(\mb S\cdot\mb E)-\frac{\gamma}{\gamma+1}(\mb S\cdot\mb v)
(\mb v\cdot\mb E)+ \right. \\
(\mb S\cdot[\mb v\times\mb B])\Biggr]+
\frac{e}{2m^2}\frac{\gamma}{\gamma+1}(\mb S\cdot[\mb
v\times\nabla])\left[ \left(g-1+\frac{1}{\gamma}\right)(\mb
S\cdot\mb B)- \right. \\ \left. (g-1)\frac{\gamma}{\gamma+1}(\mb
S\cdot\mb v)(\mb v\cdot\mb B)+
\left(g-\frac{\gamma}{\gamma+1}\right)(\mb S\cdot[\mb E\times\mb
v])\right],~~~ \gamma=\frac{1}{\sqrt{1-\mb v^2}},
\end{array} \label{eq1} \end{equation}
where $g=2\mu m/(eS)$, $\mb v$ is the velocity operator, $\gamma$
is the Lorentz factor, $Q$ is the quadrupole moment, and $\mb S$
is the spin operator. The Hermitian form of formula (1) is
obtained by the substitution ${\cal L}\rightarrow ({\cal L}+{\cal
L}^{\dag})/2$. The total Hamiltonian is given by
\begin{equation}
{\cal H}=\sqrt{m^2+\mb\pi^2}+e\Phi-({\cal L}_1+{\cal L}_2),
\label{eq2} \end{equation} where $\mb\pi\!=\!\gamma m\mb v$ is the
operator of kinetic momentum and $\Phi$ is the potential of the
electromagnetic field. We neglect the commutators of the operators
of dynamic variables.

The polarization of the electrons of the medium does not change
the form of Hamiltonian (2) if a beam contains neither electrons
nor positrons. This is attributed to the fact that the average
field acting on particles in the medium is characterized by the
electric field strength $\mb E$ and the magnetic induction $\mb
B$. However, the form of the Hamiltonian is changed if the beam
consists of electrons or positrons. There is an exchange
interaction between electrons, which is very strong.\footnote
{$^)$Recall that the exchange interaction is responsible for the
ferromagnetism.}$^)$ In the nonrelativistic case ($v\ll c$), the
magnetic field for electrons should be replaced in (2) by the
effective quasimagnetic field \cite{Bar1}
\begin{equation}
\mb B\rightarrow \mb G_e=\mb B+\mb H^{c}_{eff}+\mb H^{m}_{eff},
~~~ \mb H^{c}_{eff}=-\frac{4\pi |e|N}{mv^2}\mb P,~~~\mb
H^{m}_{eff}=\frac{2\pi |e|N} {m}(\mb P\cdot\mb n)\mb n,
\label{eq3}
\end{equation} where $N$ and $\mb P=<\mb\sigma'>$ are the
polarization density and vector (average spin), respectively, of
polarized matter electrons and $\mb n=\mb v/v$. The main
contribution to the effective quasimagnetic field, $\mb
H^{c}_{eff}$, is made by the Coulomb exchange interaction, or the
Coulomb scattering. The exchange magnetic scattering yields the
lesser contribution, $\mb H^{m}_{eff}$.

   For nonrelativistic positrons in polarized media, the effective
field with an allowance for the annihilation interaction, $\mb
H^{a}_{eff}$, is determined by
\begin{equation}
\mb G_p=\mb B+\mb H^{a}_{eff}=\mb B-\frac{\pi |e|N}{m}\mb P.
\label{eq4} \end{equation}

Formulas (3) and (4) can be represented in a more compact form by
introducing the magnetization vector (magnetic moment of a unit
volume) $\mb M$:
\begin{equation}
\mb G_e=\mb B+\frac{8\pi}{v^2}\mb M-4\pi(\mb M\cdot\mb n)\mb n,
~~~ \mb G_p=\mb B+2\pi\mb M, ~~~ \mb M=-\frac{|e|N}{2m}\mb P.
\label{eq5} \end{equation}

For isotropic magnetic materials, one can introduce a magnetic
permeability $\mu_m$:
\begin{equation}
\mb G_e=\mb B+\frac{2(\mu_m-1)}{\mu_m v^2}\mb B-
\frac{\mu_m-1}{\mu_m}(\mb B\cdot\mb n)\mb n,~~~ \mb
G_p=\frac{3\mu_m-1}{2\mu_m}\mb B. \label{eq6} \end{equation}

An appropriate expression for the Hamiltonian is expressed as
($\mb G=\mb G_e,\mb G_p$) \cite{JETP}
\begin{equation}
{\cal H}=\sqrt{m^2+\mb\pi^2}+e\Phi+\frac{e}{2m}\biggl\{g(\mb
S\cdot\mb G)+ (g-1)(\mb S\cdot[\mb E\times\mb v])\biggr\}.
\label{eq7} \end{equation}

\vskip 5mm

%\section {Equations of motion of particles and spin}
\noindent{\bfseries 3 ~~ Equations of motion of particles and
spin}

\vskip 5mm

The equation of particle motion in both polarized and unpolarized
media is the same if the Hamiltonian remains unchanged.

For electrons and positrons, the equation of particle motion is
given by \cite{JETP}
\begin{equation}
\frac{d\mb\pi}{dt}=e\mb E+e[\mb v\times\mb B]-
\frac{e}{2m}\nabla\biggl\{g(\mb S\cdot\mb G)+ (g-1)(\mb S\cdot[\mb
E\times\mb v])\biggr\}. \label{eq8} \end{equation}

For particles with arbitrary spin, the equation of spin motion
with regard to the terms quadratic in spin is given in \cite{YP}.
For nonrelativistic electrons and positrons, the equation of the
spin motion takes the form \cite{JETP}
\begin{equation}
\frac{d\mb S}{dt}=\frac{e}{2m}\biggl\{g[\mb S\times\mb G]+
(g-1)\left[\mb S\times [\mb E\times\mb v]\right]\biggr\}.
\label{eq9}
\end{equation}

The effect of the exchange interaction on the spin motion is
stronger than that on the particle motion. The equations can be
used for dia-, para-, and ferromagnetic media.\vskip 5mm

%\section {Polarization of Positronium Beams by Polarized Media}
\noindent{\bfseries 4 ~~ Polarization of positronium beams by
polarized media}

\vskip 5mm

Positronium beams can be polarized under passing through the
polarized medium. Such a possibility takes place due to a
dependence of the ortho-para-conversion (spin-conversion) rate on
the polarization of the positronium.

Positronium atoms have spin 0 (para-positronium, p-Ps) or 1
(ortho-positronium, o-Ps). Only o-Ps atoms pass through the medium
because p-Ps atoms annihilate very quickly. As a rule, the
positronium energy does not exceed several eV. In the matter, the
o-Ps lifetime can be shortened by several processes, namely, the
pick-off annihilation, the ortho-para-conversion that is the
spin-conversion, and chemical reactions \cite{Gol}.

The spin-conversion takes place in para- and ferromagnetic media,
whose molecules contain unpaired electrons. In these media, the
spin-conversion rate is generally much more than the rates of the
pick-off annihilation and other processes. As particular, in the
oxygen gas the spin-conversion rate is dozens of times larger than
the pick-off annihilation one \cite{SCSH}. The same conclusion can
be made from an analysis of experimental data on the
spin-conversion in solutions of HTMPO \cite{HTMPO}.

We consider the simplified description of the positronium
polarization process. For a more detailed description, the method
and results obtained in Ref. \cite{S} can be used.

 Let all the unpaired electrons
of the matter be polarized along the $z$-axis (Fig. 1). The spin
exchange interaction can cause changing the o-Ps spin or its
projection. This process can result in both the
ortho-para-conversion and the prompt annihilation of the o-Ps.

However, the simple analysis shows that the ortho-para-conversion
is only possible when the o-Ps spin projection onto the $z$-axis
equals either $-1$ or 0 (see Figs. 1a,b). When it equals 1,
flipping the spin is forbidden (see Fig. 1c).

The operator of the spin exchange interaction has the form
\begin{equation}
P=-J(1+4\mb s\cdot\mb s')/2, \label{eq10}\end{equation} where $\mb
s$ and $\mb s'$ are the spin operators of the o-Ps electron and
the matter electron, respectively, and $J$ is the exchange
integral that determines splitting energy levels. The operator $P$
mixes the states of the o-Ps and p-Ps.

The o-Ps with $S_z=-1$ interacting with a matter electron
($s_z=1/2$) is described by the spin wave function
$|1,-1;1/2,1/2\rangle$. As a result of simultaneous flipping the
spins of the o-Ps electron and the matter one, the o-Ps can
convert into the p-Ps with the spin wave function
$|0,0;1/2,-1/2\rangle. $ This process is characterized by the
matrix element\footnote {$^)$All the matrix elements of the
operator $\mb s\cdot\mb s'$ are given, e.g., in Ref.
\cite{LL}.}$^)$
$$\langle0,0;1/2,-1/2|P|1,-1;1/2,1/2\rangle=-J/\sqrt2.$$
Moreover, flipping the electron spins without the
o-Ps$\rightarrow$p-Ps conversion can also occur. The o-Ps spin
becomes zero. The matrix element characterizing this process
equals
$$\begin{array}{c} \langle1,0;1/2,-1/2|P|1,-1;1/2,1/2\rangle=
-J/\sqrt2. \end{array}$$

Analogous processes take place for the o-Ps with $S_z=0$. The spin
exchange interaction between the o-Ps electron and the matter one
with and without the ortho-para-conversion is described by the
matrix elements
$$\begin{array}{c} \langle0,0;1/2,1/2|P|1,0;1/2,1/2\rangle\!=\!-J/2 ~~
 {\rm and} ~~
 \langle1,1;1/2,-1/2|P|1,0;1/2,1/2\rangle\!=\!
-J/\sqrt2,  \end{array}$$ respectively. Coupled electrons do not
contribute to the spin-conversion because matrix elements
characterizing the exchange interaction between the o-Ps and the
pair of coupled electrons are zero.

For the o-Ps with $S_z=1$, all the corresponding matrix elements
are zero. Therefore, the majority of o-Ps atoms passes through the
medium without the annihilation. The pick-off annihilation is
possible for the o-Ps atoms with any spin projection.

The o-Ps lifetime is changed by a magnetic field into the medium.
This field can be strong enough. As is known, such a field does
not change the lifetime of the o-Ps with $S_z=\pm1$ and shortens
the lifetime of the o-Ps with $S_z=0$ (see Ref. \cite{Gol}). This
circumstance accelerates the process of polarization of the o-Ps
beam. However, the evaluation shows such an acceleration is not
very significant. For example, the pick-off annihilation shortens
the o-Ps lifetime more greatly than the magnetic field in
ferromagnetic media.

\vskip 5mm

%\section {Discussion and conclusions}
\noindent{\bfseries 5 ~~ Discussion and conclusions}

\vskip 5mm

Magnetic crystals can be effectively used for the rotation of the
polarization vector of particles. Even for neutrons, whose
magnetic moment is relatively small, for $B\sim 1$ T, the angle of
rotation of the polarization vector per unit length is of the
order of $\Delta\Phi/\Delta l\sim (c/v)\times 10^{-3}$ rad/cm.

The rotation of the polarization vector in magnetic crystals
reaches especially large values for nonrelativistic electrons. It
follows from (6) and (7) that the angular velocity of the spin
precession of nonrelativistic electrons is increased by the factor
of $(c/v)^2$ due to the exchange interaction. For $B\sim 1$ T, we
have $\Delta\Phi/\Delta l\sim (c/v)^3\times 1$ rad/cm in order of
magnitude. In particular, when $v/c\sim 0.1$, we have
$\Delta\Phi/\Delta l\sim 10^{3}$ rad/cm.

The use of magnetic crystals may also be sufficiently effective
for the rotation of the polarization vector of relativistic
electrons.

The Stern--Gerlach force, which splits beams according to the
polarization of particles, is considerably increased. However, the
use of polarized media for splitting electron beams according to
the polarization is seriously hampered by the small value of the
energy of interaction between the spin of electrons and a
quasimagnetic field ${\cal W}^{(s)}$ (of the order of 1 eV or
less) and a multiple scattering that increases the transverse
energy of electrons. If the transverse energy of electrons is
greater than $|{\cal W}^{(s)}|$, splitting the beam according to
the polarization becomes impossible.

The formulas obtained in this work are also valid for beams of
polarized nuclei.

Media with polarized electrons can be used for the polarization of
positronium beams. The spin direction of positroniums coincides
with the spin direction of polarized electrons.

The considered effect of the polarization of positronium beams is
more important for para- and ferromagnetic media without
conductivity electrons, e.g., ferrites. The spin exchange
interaction of o-Ps atoms with conductivity electrons leads to the
spin-conversion that does not depend on the o-Ps polarization. As
a result, a presence of conductivity electrons can cause a strong
background making the effect unobservable.

It is important that the intensity of the positronium beam passing
through two samples magnetized in different directions, depends on
the angle $\phi$ between the magnetization axes. This effect is
similar to passing the unpolarized light through two polarizers
when their axes do not coincide.

Polarized positronium beams can be used for an investigation of
magnetic media. The investigation can include monitoring the
process of magnetization and determining the magnetic structure of
materials. \vspace{3mm}

The work is supported by the grant of the Belarusian Republican
Foundation for Fundamental Research No. $\Phi$03-242.

%%%%%%%%%%%%%%%%%%%%%%%%%%%%%%%%%%%%%%%%%%%%%%%%%%
\newpage
\begin{wrapfigure}{R}{15.0029cm}

\mbox{\epsfig{figure=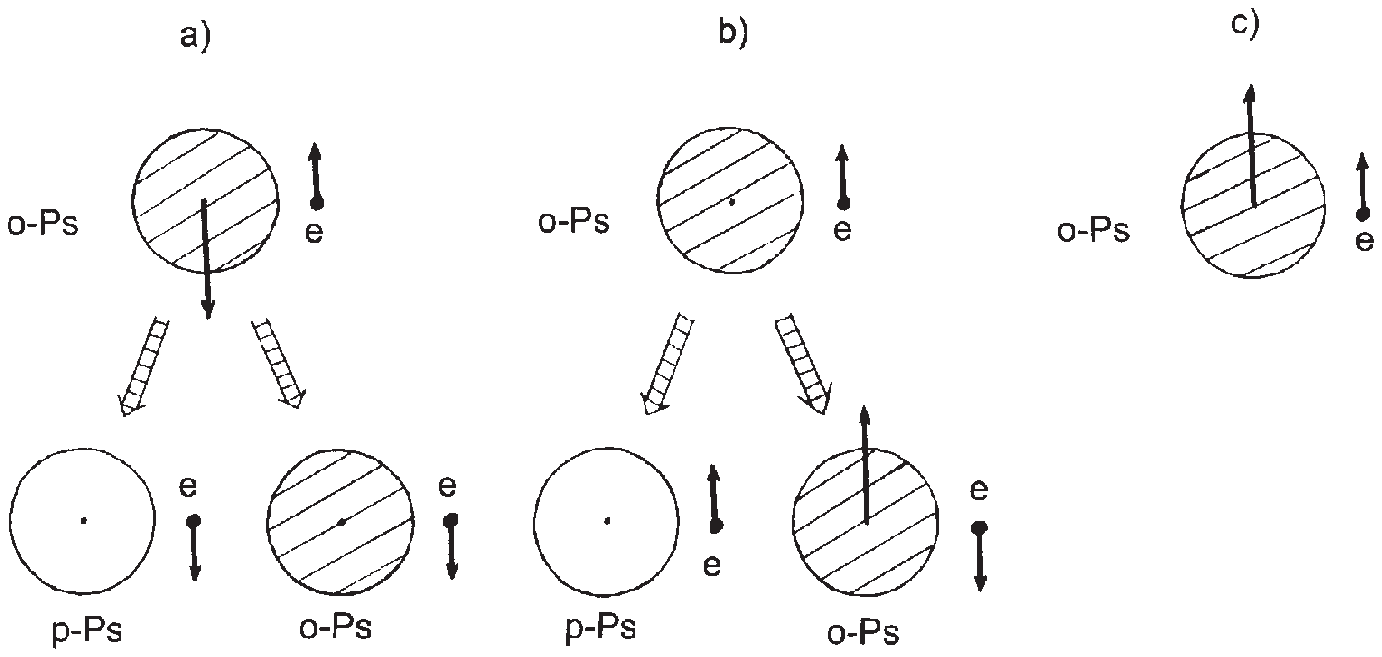,width=15.0029cm,height=7.1289cm}}

{\bf Figure 1.} Spin-conversion of the o-Ps: a) the o-Ps (hatched)
with the spin projection $S_z=-1$ can convert either into the p-Ps
(white) or into the o-Ps with $S_z=0$; b) the o-Ps with $S_z=0$
can convert either into the p-Ps with $S_z=-1$ or into the o-Ps
with $S_z=1$; c) the o-Ps with $S_z=1$ cannot convert.

\end{wrapfigure}

%%%%%%%%%%%%%%%%%%%%%%%%%%%%%
\end{document}